\documentclass[12pt]{article}
\usepackage{amsmath,bm,amsfonts,color,amsthm, amssymb}
\usepackage[margin=2.5cm]{geometry}
\usepackage{graphicx}
\usepackage{natbib}
\usepackage{enumitem}
\usepackage{subcaption,siunitx,booktabs}
\usepackage[citecolor=blue,urlcolor=blue,allcolors=blue]{hyperref}
\usepackage{arydshln}
\usepackage{chngcntr}
\usepackage[utf8x]{inputenc}

\usepackage{titlesec}
\titlelabel{\thetitle.\quad}

\usepackage{tikz}
\usepackage{transparent}
\usepackage[toc,page]{appendix}

\usetikzlibrary{positioning}
\usetikzlibrary{shadows}
\usepgflibrary{shapes}

\usepackage{mathtools}

\usepackage{apptools}
\AtAppendix{\counterwithin{lemma}{section}}

\AtAppendix{\counterwithin{proposition}{section}}

\usepackage{titlesec}

\titleformat*{\section}{\Large\bfseries}
\titleformat*{\subsection}{\large\bfseries}
\titleformat*{\subsubsection}{\large\bfseries}
\titleformat*{\paragraph}{\large\bfseries}
\titleformat*{\subparagraph}{\large\bfseries}

\usepackage{titlesec}
\titlelabel{\thetitle\quad}

\theoremstyle{definition}

\numberwithin{mytheorem}{section}
\usepackage{setspace}

\newtheorem{myremark}{Remark}
\title{Bayesian Estimation of Variance under Fine Stratification via Mean-Variance Smoothing}

\author{Sepideh Mosaferi \footnote{Florida International University. Corresponding author (E-mail: \href{mailto:smosaferi@umass.edu}{smosafer@fiu.edu})} and Shonosuke Sugasawa \footnote{Faculty of Economics, Keio University}}

\date{}

\begin{document}

\maketitle

\doublespacing

\begin{abstract}
Fine stratification survey is useful in many applications as its point estimator is unbiased, but the variance estimator under the design cannot be easily obtained, particularly when the sample size per stratum is as small as one unit. One common practice to overcome this difficulty is to collapse strata in pairs to create \textit{pseudo-strata} and then estimate the variance. The estimator of variance achieved is not design-unbiased, and the positive bias increases as the population means of the paired pseudo-strata become more variant. The resulting confidence intervals can be unnecessarily large. In this paper, we propose a new Bayesian estimator for variance which does not rely on collapsing strata, unlike the previous methods given in the literature. We employ the penalized spline method for smoothing the mean and variance together in a nonparametric way. Furthermore, we make comparisons with the earlier work of Breidt et al. (2016). Throughout multiple simulation studies and an illustration using data from the National Survey of Family Growth (NSFG), we demonstrate the favorable performance of our methodology.

\vspace{0.25cm}

\end{abstract}

\clearpage

\section*{Statement of Significance} The article demonstrates a new Bayesian approach to estimate the variance for the commonly used fine stratification designs and uses the penalized spline method to smooth the mean and variance simultaneously. The method does not rely on collapsing strata, unlike previous works in the literature. We provide two simulation studies and an illustration using real data set to support our work and make comparisons.

\section{Introduction} \label{sec:introduction}

The process of independently selecting a very small number of units -- often one or two primary sampling units (PSUs) -- per disjoint countable subpopulations (so-called strata) is called fine stratification.
Surveys such as the Survey of Income and Program Participation of the U.S. Census Bureau \citep{Westat2001} and the U.S. Department of Agriculture's National Resources Inventory \citep{nusser1997national, Breidt2002} use a multi-stage two-per-stratum design and a stratified two-stage area sampling design, respectively. 
The Canadian Health Measures Survey \citep{harold2009variance} with a three-stage sample design is also based on the small number of PSUs per stratum.
The Current Population Survey \citep{USCB2006} and the National Crime Victimization Survey (\cite{Lohr1999}, Section 7.6) both conducted by the U.S. Census Bureau, and the National Survey of Family Growth \citep{lepkowski20102006} conducted by the University of Michigan's Institute for Social Research under contract to the U.S. National Center for Health Statistics are based on multi-stage one-per-stratum designs.

Despite the popularity of the design, which allows the usual \cite{horvitz1952generalization} estimator for the parameter of interest, such as population mean or total, to be unbiased and efficient, it is also recognized to have a number of caveats. In particular, the variance under the design for one PSU per stratum does not exist and for two or three PSUs per stratum contains a large amount of variation. 
A traditional method to estimate the variance of an unknown parameter under fine stratification is to collapse the adjacent strata to create pseudo-strata with more PSUs and then calculate the variance. This method was first introduced by \cite{hansen1953sample}, but it often causes an overestimation in the variance of the estimator. 

In this paper, we propose a Bayesian approach as an alternative method to estimate variance under fine stratification. The advantages of the proposed Bayesian method are as follows: (i) it does not rely on collapsing strata, (ii) it models the mean and variance of strata simultaneously, which resulting in significantly improved performance compared with the existing methods, and (iii) it can be easily implemented and provides measure of uncertainty in addition to point estimates.

\subsection{Primary Sampling Unit} \label{subsec:PSU}

In applications, the types of units that are called PSUs depend on the survey. The PSU is the first stage sampling unit. It could be the final unit of interest, but it is usually a cluster of some forms such as a school, hospital, or business unit; in area probability samples it typically consists of counties, census tracts, block groups, or local administrative units. To retain equal probability of selection, the PSUs are usually selected with probabilities proportional to their population sizes, with a second stage of selection sampling an equal number of SSUs, but unequal probabilities of inclusion can be undertaken for other reasons (e.g., non-response adjustment). For example, in the 1998 Survey of Mental Health Organizations (SMHO), hospitals are PSUs. In surveying schools, the schools are PSUs (see, \cite{valliant2013practical} Section 10.3 for some further examples).   
In the simulation studies given in this article, we consider sampling units as our PSUs of interest.

In area samples, there are enough strata to select only one or two PSUs per stratum. 
Selecting one PSU per stratum allows more control over the achieved distribution of the sample than selecting two or more PSUs, but creates some variance estimation problems that we address in this paper.

\subsection{Some Related Works} \label{subsec:literature}

Several alternatives have been proposed in the literature to overcome the limitations of collapsed strata variance estimators. 
\cite{hansen1953sample} and \cite{isaki1983variance} proposed to use some auxiliary variables well-correlated with the expected values of the mean of the strata to reduce the bias of collapsed strata variance estimator. \cite{harold2009variance} proposed an approach based on variance components from different sampling stages and applied it to the Canadian Health Measures Survey.
\cite{mosaferiempirical} proposed an empirical Bayesian estimator as well as a constrained empirical Bayesian estimator for the variance of one unit per stratum sample design. The author also compared one PSU per stratum design with two PSUs per stratum design and pointed out some of the instabilities of the proposed estimators due to the method of moment parameter estimation.

Recently, \cite{breidt2016nonparametric} proposed a nonparametric method that replaces a collapsed stratum estimator by a kernel-weighted stratum neighborhood estimator, which uses the deviations from a fitted mean function to estimate the variance. The underlying assumption in their paper is that stratum means (or totals) vary smoothly with an auxiliary variable. Their proposed method still relies on collapsing strata.

Regarding Bayesian approaches under informative sampling, \cite{wang2018approximate} introduced an approximate Bayesian method under informative sampling and it was applied to regression estimation \citep{sugasawa2022approximate}.
In addition, \cite{zhao2020bayesian} employed Bayesian empirical likelihood under informative sampling.
However, there are no existing approaches tailored for the joint estimation of means and variances.

\subsection{Organization of Paper} \label{subsec:organization}

The remainder of the paper is organized as follows. In Section \ref{sec:variance}, we provide some notation and review the variance of fine stratification based on the collapsed strata method and the kernel-based method. In Section \ref{sec:HB}, we propose the setup of our Bayesian model for one PSU and multiple PSUs per stratum, and we explain how they could be used in practice. 
Section \ref{sec:simulation} demonstrates our methodology and its performance throughout several simulation studies. Furthermore, in Section \ref{sec:application}, we apply our method to a real data set from NSFG. Some conclusions and directions for future research are given in Section \ref{sec:discuss}. Supplementary material contains details of the derivations and additional simulation results. 

%All the \texttt{R} code implementing the proposed methodology are provided in a zip file.

%available at Github repository \url{https://github.com/SepidehMosaferi/Bayesian-Estimator-for-Variance}.

\section{Variance of Fine Stratification} \label{sec:variance}

In stratified sampling designs, we partition the target population into a disjoint finite number of strata where the units per each stratum are homogeneous and share at least one common characteristic, such as location, income, education, or gender. This partitioning of variance in combination with the distribution of the strata in the population being known causes the variance of the parameter of interest to be reduced relative to a simple random sample. Also, stratification allows for a flexible method of sample selection per stratum, such as simple random sampling with or without replacement and systematic sampling as strata are independent. In this paper, we focus on the particular case of fine stratification design where the sample size per stratum is as small as one PSU or two PSUs selected without replacement. 
Our target parameter is the population mean, and our goal is to estimate the variance of an estimated mean. Often, this variance brings about complications since it is not estimable when the sample size per stratum is as small as one unit. Throughout the rest of this section, we provide some notation and explain the traditional collapsed strata variance estimator and kernel-based variance method.

\subsection{Notation and Estimation} \label{subsec:notation}

Consider a finite population with $N$ units and index set $U=\{1,2,...,N\}$, where they are divided into $H$ disjoint strata such as $U=\cup_{h=1}^{H}U_h$ each with the size of $N_h$ for $h=1,2,...,H$ such that $N=\sum_{h=1}^{H}N_h$.
Let $y_{hj}$ denote the value of the $j$-th individual nested in the $h$-th stratum such that $j=1,...,N_h$. 
Additionally, assume that the value of $y_{hj}$ is observed without an error. 
Randomly and without replacement, we select a sample with a size of $n_h$ from the $h$-th stratum, where $n_h \ge 1$, and the total sample size is $n=\sum_{h=1}^{H}n_h$. 

Let $u_{h}$ denote the selected sample with the size of $n_h$ from stratum $h$ such that the total sample is $u=u_1 \cup u_2 \cup ... \cup u_H$, where $u_h \subset U_h$. Also, let $\pi_j=\mathbb{P}\{j \in u_{h} \}$ be the first-order inclusion probability and $\pi_{jk}=\mathbb{P}\{ j,k \in u_h \}$ be the second-order inclusion probability. The finite population mean under stratified design is $\bar{y}= N^{-1}\sum_{h=1}^{H} \sum_{j=1}^{N_h} y_{hj}$.
The \cite{horvitz1952generalization} unbiased estimator of the finite population mean under stratified design is
\begin{equation} \label{eq:HTestimator}
\bar{y}_{\text{HT}}=\frac{1}{N} \sum_{h=1}^{H} \sum_{j=1}^{n_h} \frac{y_{hj}}{\pi_j},
\end{equation}
with the variance of 
\begin{align} \label{eq:true_var}
V(\bar{y}_{\text{HT}}) & = N^{-2} \sum_{h=1}^{H} S^2_h \qquad \text{where} \nonumber\\  
S^2_h & =\mathop{\sum \sum}_{(j,k )\in U_h} \Delta_{jk} (y_{hj}/\pi_j) (y_{hk}/\pi_k) \quad \text{with}  \quad
\Delta_{jk}=\pi_{jk}-\pi_j \pi_k.
\end{align}
An unbiased estimator of the variance of $\bar{y}_{\text{HT}}$ given in (\ref{eq:true_var}) is
\begin{equation} \label{eq:varStratified} 
var(\bar{y}_{\text{HT}})=\frac{1}{N^2} \sum_{h=1}^{H} s^2_h \quad \text{where} \quad s^2_h=\mathop{\sum \sum}_{(j,k )\in u_h} \frac{\Delta_{jk}}{\pi_{jk}} \frac{y_{hj}}{\pi_j} \frac{y_{hk}}{\pi_k}.
\end{equation}

In many applications, the sampling units $u_h$'s are the primary stage units.
In this paper, we only concentrate on the single-stage sampling and assume that we have a stratified design (see \cite{valliant2013practical} chapter 3 for the definition of this design). This design could be easily generalized to more complicated survey designs such as multi-stage two-per-stratum design or multiple stages of selection within strata.

\subsection{Collapsed Strata Variance Estimator} \label{subsec:collapsed}

When there is only one PSU per stratum, the variance estimator given in equation (\ref{eq:varStratified}) cannot be obtained, since $\pi_{jk}=0$. Instead, a collapsed variance estimator can be used, where we create larger strata by often pairing the adjacent two strata to have more sampling units per new pseudo-stratum. Pseudo-strata can be created by using spatial locations or strata size measures, which often come from the population frame, and should be done before the process of sample selection. For simplicity, assume that the total number of strata ``$H$" is an even number. If $\{x_h\}$ denotes the collapsing index for $h=1,...,H$, then we can create successive pseudo-strata by sorting strata based on $x_h$. 

The collapsed strata variance estimator is
\begin{equation} \label{eq:varCollapsed}
var_{\text{Coll}}=\frac{1}{N^2} \sum_{h=1}^{H} s^2_{\text{Coll-h}} \quad \text{where} \quad s^2_{\text{Coll-h}}= \frac{1}{2} \Big(\sum_{j \in u_h} \frac{y_{hj}}{\pi_j} - \sum_{l=1}^{H} c_{l(h)} \sum_{k \in u_l} \frac{y_{lk}}{\pi_k}  \Big)^2,
\end{equation} 
and
\begin{equation*}
c_{l(h)}  =\left\{\begin{array}{cc}
1 & \quad \text{$h \neq l$ belong to the same collapsed stratum;} \\
0 & \quad \text{otherwise.}\end{array}\right.
\end{equation*}
For the collapsed strata variance estimator given in (\ref{eq:varCollapsed}) see also \cite{breidt2016nonparametric}, \cite{sarndal2003model} p. 109, and \cite{wolter2007introduction} sec. 2.5. 

The variance estimator in (\ref{eq:varCollapsed}) is design-biased, and its bias with respect to the design is 
\begin{equation} \label{eq:bias}
\text{Bias}(var_{\text{Coll}})=\frac{1}{2 N^2}\sum_{h=1}^{H}\Big(\sum_{j \in U_h}y_{hj}- \sum_{l=1}^{H} c_{l(h)} \sum_{k \in U_l}y_{lk}\Big)^2.
\end{equation}
The bias term is positive, and it becomes small if $\sum_{j \in U_h}y_{hj} \approx \sum_{k \in U_l}y_{lk}$ for $c_{l(h)}=1$. To further simplify the idea, assume that $N_h=N/H$ and that only one unit is selected per stratum. Then, the bias term is proportional to $2H^{-2} \sum_{(h \neq l)=1}^{H} (\bar{y}_h- \bar{y}_l)^2$ under srswor selection, where $\bar{y}_h=N_h^{-1}\sum_{j \in U_h}y_{hj}$ and $\bar{y}_l$ can be defined in a similar way. The bias term is of the order $O(H^{-1})$ as $N \rightarrow \infty$, which suggests a strategy on how we can group the strata in pairs to reduce the amount of bias with respect to the characteristic of interest to minimize the difference $|\bar{y}_h-\bar{y}_l|$.     

We expect that as the number of strata increases, the amount of bias decreases. When $``H"$ is an odd number, we can create a single pseudo-stratum by combining three original strata (say) and the rest of pseudo-strata by combining two original strata. Then, we use the same methodology described in this paper to estimate the variance. \cite{rust1987strategies} compared variance estimators by collapsing two, three or more strata. They showed that increasing the number of collapsed strata results in a more biased variance estimator. Hence, collapsing strata in pairs seems to be a good strategy.
Also, for the kernel-based variance estimator proposed by \cite{breidt2016nonparametric}, the strata should be paired in advance, particularly for one PSU case per stratum; otherwise, the variance term is not identifiable. Unlike previous works in the literature, our proposed estimator in Section \ref{sec:HB} does not rely on collapsing strata.

\subsubsection{Creating Pseudo-Strata} \label{subsec:Pseudo-Strata}

Often, strata should be combined based on the characteristics of strata such as size, location, temporal structure, or stratum total (based on a frame variable) prior to the sample selection. Let $x_h$ denote the collapsing index, which is known for every stratum $h=1,...,H$. Then the collapsing function $c_{l(h)}$ is constructed by sorting the strata along the known values of $x_h$ to form successive pairs of strata.
The pairing strata should be set once and should not depend on the results of the selected samples. If $\bar{y}_h$'s are correlated with $x_h$'s, then the pairing is effective and the bias given in expression (\ref{eq:bias}) is small. 
For example, if population size or spatial location were used to form strata, then two strata with similar sizes of PSUs could be combined. Combining strata based on the characteristics of the selected sample can lead to negatively biased variance estimates. 

\subsection{Kernel-Based Variance Estimator} \label{subsec:Kernelvar}

\noindent The nonparametric kernel-based variance estimator given by \cite{breidt2016nonparametric} is as follows
\begin{equation} \label{eq:varKer}
var_{\text{Ker}}=\frac{1}{N^2} \sum_{h=1}^{H} s^2_{\text{Ker-h}} \quad \text{where} \quad 
s^2_{\text{Ker-h}}= \frac{1}{C_d}\Big(\sum_{j \in u_h} \frac{y_{hj}}{\pi_j}- \sum_{l=1}^{H} d_{l(h)} \sum_{k \in u_l} \frac{y_{lk}}{\pi_k} \Big)^2,
\end{equation}  
where the binary function $d_{l(h)}$ given in the collapsed strata variance (\ref{eq:varCollapsed}) is replaced by the following kernel weight 
\begin{equation*}
d_{l(h)}=K\Big(\frac{x_h-x_l}{b}\Big)/\sum_{l=1}^{H}K\Big(\frac{x_h-x_l}{b}\Big),
\end{equation*}
such that $K(.)$ is a symmetric bounded kernel function and $b$ is the bandwidth parameter. 

The nonrandom normalizing constant $C_d$ is
\begin{equation*}
C_d=\frac{1}{H} \sum_{h=1}^{H} \Big(1-2d_{h(h)}+\sum_{l=1}^{H}d_{l(h)}^2\Big),
\end{equation*}
which only depends on the kernel weights. The Epanechnikov kernel $K(u)= 0.75 (1-u^2) 1_{\{|u| \leq 1 \}}$ is used for the kernel weight $d_{l(h)}$. The kernel estimator $s^2_{\text{Ker-h}}$ given in (\ref{eq:varKer}) is biased; see Section 3 of \cite{breidt2016nonparametric}. 

\begin{myremark}
The estimator (\ref{eq:varKer}) is biased, therefore, $C_d$ should be selected in such a way as to reduce the bias caused by $s^2_h$ in (\ref{eq:varStratified}) if the $s^2_h$'s are constant throughout the strata.
\end{myremark}

\begin{myremark}
For the kernel-based variance estimator, the bandwidth should tend to zero ($b \rightarrow 0$) and satisfy the condition of $Hb^2 \rightarrow \infty$ based on A6 given in \cite{breidt2016nonparametric}. 
In their guidepost,  they recommend to pick the bandwidth between $(1/H,2/H)$. In this paper, we assume $b=(1/H+2/H)/2$ for simulations and the illustrative analysis of the NSFG data set.
\end{myremark}

Some of the main disadvantages of (\ref{eq:varKer}) are as follows: (1) using a symmetric kernel when the support of the distribution is not on the entire real line, particularly in applications related to income, salary, etc. causes weights to be assigned outside of the domain of the observations, resulting in boundary bias (see \cite{bouezmarni2005consistency}), (2) there could be potential cases where $C_d$ is equal to zero, which makes $var_{\text{Ker}}$ undefinable, (3) the selection of bandwidth should be done carefully, (4) $d_{l(h)}$ depends on the covariate $x_h-x_l$, which may bring difficulties in some boundary cases, and (5) the method still relies on collapsing strata.

\section{Bayesian Variance Estimator} \label{sec:HB}

In this section, we describe our Bayesian estimator, where we first define two unknown functions for the mean and variance per each stratum. Then, we intend to smooth these functions simultaneously across all strata and eventually estimate the unknown parameters as well as the variance.

\subsection{Design with One PSU per Stratum}\label{sec:1PSU}

First, we consider a simple case with $n_h=1$ (single unit per each stratum). 
We assume the following model for $y_h$:
\begin{equation}\label{eq:model-PSU1}
y_h\sim N(m(x_h), s^2(x_h)), \ \  \ \ h=1,\ldots,H,
\end{equation}
where $m(x_h)$ and $s^2(x_h)$ are unknown functions of mean and variance per each stratum, respectively. We assume $x_h$ is the collapsing index or any other available covariate from the
$h$-th stratum.

For modeling the unknown functions, we employ the penalized spline method \citep[e.g.][]{ruppert2002selecting,opsomer2008non} described as follows:

\begin{align} 
m(x_h; \beta) & =\beta_0+\beta_1 x_h+\cdots + \beta_q x^q_h + \sum_{l=1}^L \beta_{q+l}(x_h-\kappa_l)_+^q, \quad \text{and} \label{eq:m_function} \\
\log s^2(x_h; \gamma) & =\gamma_0+\gamma_1 x_h+\cdots +\gamma_q x^q_h + \sum_{l=1}^L \gamma_{q+l}(x_h-\kappa_l)_+^q. \label{eq:s2_function}
\end{align}

In equations (\ref{eq:m_function}) and (\ref{eq:s2_function}), $(\beta_0,\beta_1,\ldots,\beta_q, \beta_{q+1}, ... , \beta_{q+L})$ and $(\gamma_0,\gamma_1,\ldots,\gamma_q, \gamma_{q+1}, ... , \gamma_{q+L})$ are unknown coefficients. Furthermore, $(x)_+^q=\max(x, 0)^q$, and $``q"$ is the degree of polynomial. 
Here, we assume $\kappa_1,\ldots,\kappa_L$ are knots, and $L$ is the number of knots specified by a user. 
When the knots are spread over the entire space of $x$, the spline function is sufficiently flexible even for small $q$ such as $q=2$ or $3$. 
Although we can allow different values of $q$ and $L$ for $m(x; \beta)$ and $s^2(x; \gamma)$, we use the same values for simplicity. 
In particular, we assume $q=2$ and $L=7$ in our simulations and NSFG example. 

The model (\ref{eq:model-PSU1}) allows us to vary the mean and variance parameters according to the collapsing index $``x"$, so we call the method mean-variance smoothing.  
To prevent over-fitting of the spline functions, we treat $\beta_{q+1},\ldots,\beta_{q+L}$ as random effects by assuming 
\begin{equation*}
\beta_{q+l} \overset{\text{ind}}{\sim} N(0, \tau_\beta^2), \quad \text{for} \quad l=1,\ldots,L,
\end{equation*}
where $\tau_\beta^2$ is an unknown variance parameter. 
Similarly, we assume $\gamma_{q+1},\ldots,\gamma_{q+L}$ as random effects such that
\begin{equation*}
\gamma_{q+l} \overset{\text{ind}}{\sim} N(0, \tau_\gamma^2), \quad \text{for} \quad l=1,\ldots,L,    
\end{equation*}
with the unknown parameter $\tau_\gamma^2$.

As in the model (\ref{eq:model-PSU1}), Bayesian modeling of unknown mean and variance functions has been considered in the literature by \cite{nott2006semiparametric} and \cite{pati2014bayesian}.
Here, we incorporate the information of the sampling probability $\pi_h$ in obtaining the posterior distribution.
To this end, we consider the following joint posterior: 
\begin{equation}\label{HB-pos}
\pi(\boldsymbol{\Theta})\prod_{h=1}^H \phi(y_h; m(x_h; \beta), s^2(x_h; \gamma))^{\tilde{w}_h}\prod_{l=1}^L \phi(\beta_{q+l}; 0, \tau_\beta^2)\phi(\gamma_{q+l}; 0, \tau_\gamma^2),
\end{equation}
where $\tilde{w}_h=\pi_h^{-1} /H^{-1}\sum_{h=1}^H \pi_h^{-1}$ is the normalized weight, and $\phi(\cdot; \mu, \sigma^2)$ is the density function of normal distribution with mean $\mu$ and variance $\sigma^2$. 

The detailed steps related to equation (\ref{HB-pos}) are given in Appendix C from the supplementary material.
While the posterior (\ref{HB-pos}) is called the general posterior \citep[e.g.][]{bissiri2016general}, its uncertainty could be adequately controlled by using the normalized weight satisfying $H^{-1}\sum_{h=1}^H \tilde{w}_h=1$. The presence of $\tilde{w}_h$'s allows the results to have more stability (see Appendix D from the supplementary material for a relevant simulation study).

Furthermore, $\pi(\boldsymbol{\Theta})$ in (\ref{HB-pos}) corresponds to the prior distributions of the unknown parameters $\boldsymbol{\Theta}=(\beta_0,\ldots,\beta_q, \gamma_0,\ldots,\gamma_q, \tau_\beta^2, \tau_\gamma^2)^\top$.
Note that the log-likelihood of expression (\ref{HB-pos}) can be expressed as 
\begin{equation*}
\sum_{h=1}^H \tilde{w}_h \log \phi(y_h; m(x_h; \beta), s^2(x_h; \gamma)), 
\end{equation*}
which is equivalent to the well-known pseudo-likelihood or weighted likelihood under informative sampling \citep[e.g.][]{godambe1986parameters,pfeffermann1993role}.

In order to estimate the variance of the HT estimator $\bar{y}_{\text{HT}}$, we use a hybrid method where design-based weights should be taken into account. 
Based on the posterior samples of $\gamma$, the posterior samples of $s^2(x_h;\gamma)$ can be obtained. Thus, the posterior mean can be used as its relevant point estimate. 
To this end, the Bayesian variance estimator for the HT estimator can be obtained as
\begin{equation*} \label{eq:Bayesvar}
 var_{\text{Bayes}}= \frac{1}{N^2} \sum_{h=1}^{H} w^2_h \hat{s}^2(x_h; \gamma), 
\end{equation*}
where $w_h \propto 1/\pi_h$ is a related weight based on the sampling design.

\cite{zheng2003penalized} attempted a different approach, where they modeled the conditional mean $E(y_j|\pi_j)$ by penalized splines to deal with inefficiencies in weighted estimators and studied alternatives to the HT estimator. They eventually made comparisons with the HT estimator. 
Note that we have modeled $y_j|\mathbb{I}\{j \in u_h\}$ in equation (\ref{eq:model-PSU1}), where $j=h$ notationally as $n_h=1$ here. This modeling framework could potentially address any kinds of  estimators based on the fine stratification design. 
%This differs from the formulation given in \cite{zheng2003penalized}. 
%Thus, in order to estimate the variance of the HT estimator, we consider the posterior variance for the mean. 

\subsubsection{Prior Specification and Posterior Computation}

In this section, we define prior distributions for the vector of unknown parameters $\boldsymbol{\Theta}$.
Regarding the regression coefficients for mean and variance, we assign $\beta_k  \sim N(0, S_\beta)$ and $\gamma_k  \sim N(0, S_{\gamma})$, independently for $k=1,\ldots,p$.
Here $S_{\beta}$ and $S_{\gamma}$ are fixed hyper-parameters.
Additionally, for the variance parameters, we assign $\tau_\beta^2 \sim {\rm IG}(a_{\beta}, b_{\beta})$ and $ \tau_\gamma^2  \sim {\rm IG}(a_{\gamma}, b_{\gamma})$, 
where $a_{\beta}, b_{\beta}$, $a_{\gamma}$, and $b_{\gamma}$ are fixed hyperparameters.
Here, ${\rm IG}(a,b)$ denotes the inverse-gamma distribution with the density proportional to $x^{a-1}\exp(-b/x)$ for $x>0$. 

As a default choice, we use $S_{\beta}=S_{\gamma}=100$ and  $a_{\beta}=b_{\beta}=a_{\gamma}=b_{\gamma}=1$, which results in diffuse priors on the regression coefficients and weakly informative priors on the variance parameters.  
%\textcolor{blue}{[[Shonosuke: Maybe you need to give some brief reasons for these choices?! Do we need to consider any sensitivity analysis?]]}
The joint posterior given in (\ref{HB-pos}) can be approximated by a Markov Chain Monte Carlo (MCMC) algorithm, in particular Metropolis-within-Gibbs sampling.
The details of the derivations and sampling steps from the full conditional distributions are provided in the supplementary material.

\subsection{Design with Multiple PSUs per Stratum}\label{sec:multi-PSU}
The hierarchical model in the previous section can be extended to a situation with multiple PSUs. 
Let $n_h$ be the sample size from $h$-th stratum. 
Then, we consider the following model: 

\begin{equation} \label{eq:N_multiple}
\boldsymbol{y}_h \sim N_{n_h}(m(x_h) \boldsymbol{1}_{n_h}, s^2(x_h) \boldsymbol{R}_h(\rho)), \quad h=1,\ldots,H,
\end{equation}
where $\boldsymbol{y}_h=(y_{h1},\ldots,y_{hn_h})^\top$, and $\boldsymbol{1}_n$ is an $n$-dimensional vector of $1$'s.
Additionally, $\boldsymbol{R}_h(\rho)$ in expression (\ref{eq:N_multiple}) is an $n_h\times n_h$ matrix whose $(i,j)$-element is $\rho^{I(i\neq j)}$; i.e.

\begin{equation*}
\boldsymbol{R}_h(\rho)=   \begin{pmatrix}
    1 & \rho & \rho & \dots  & \rho \\
    \rho & 1 & \rho & \dots  & \rho \\
    \vdots & \vdots & \vdots & \ddots & \vdots \\
    \rho & \rho & \rho & \dots  & 1
\end{pmatrix},
\end{equation*}
where $\rho\in (-1,1)$ is an unknown correlation parameter.
Here, $m(x_h)$ and $s^2(x_h)$ are scalar quantities as a function of the collapsing index $x_h$. 

For the mean function $m(x)$ and the variance function $s^2(x)$, we employ the same P-spline formulation defined for the single PSU case discussed in Section~\ref{sec:1PSU}. 
We define $\tilde{w}_{hj}$ as follows:
\begin{equation*}
    \tilde{w}_{hj}=\frac{\pi_{hj}^{-1}}{(\sum_{h=1}^{H}\sum_{j=1}^{n_h}\pi_{hj}^{-1})/\sum_{h=1}^{H}n_h},
\end{equation*}
where $\sum_{h=1}^{H}\sum_{j=1}^{n_h} \tilde{w}_{hj}/\sum_{h=1}^{H}n_h=1$. 
To define the posterior, we need to define the power likelihood as used in (\ref{HB-pos}).

Let $\boldsymbol{z}_h=\boldsymbol{R}_h(\rho)^{-1/2}\{\boldsymbol{y}_h-m(x_h,\beta)\boldsymbol{1}_{n_h}\}$, which follows $N(\boldsymbol{0}_{n_h}, s(x_h,\gamma) \boldsymbol{I}_{n_h})$ under (\ref{eq:N_multiple}).
Then, the power likelihood for $\boldsymbol{z}_h=(z_{h1},\ldots,z_{hn_h})$ can be defined as 
$$
\prod_{j=1}^{n_h} \phi(z_{hj}; 0, s(x_h,\gamma))^{\tilde{w}_{hj}}
\propto 
s(x_h,\gamma)^{-\sum_{j=1}^{n_h}\tilde{w}_{hj}/n_h} \exp\left\{-\frac{\boldsymbol{z}_h^\top \boldsymbol{W}_h\boldsymbol{z}_h}{2s(x_h,\gamma)}\right\},
$$
where $\boldsymbol{W}_h={\rm diag}(\tilde{w}_{h1},\ldots, \tilde{w}_{hn_h})$. 
By changing the variable, the power likelihood of $\boldsymbol{y}_h$ is proportional to 
\begin{align*}
&L_w(\boldsymbol{y}_h; \beta, \gamma, \rho)\\
&=|\boldsymbol{R}_h(\rho)|^{-1/2}s(x_h,\gamma)^{ -\sum_{j=1}^{n_h}\tilde{w}_{hj}/n_h} \exp\left[-\frac{\boldsymbol{r}_h(\boldsymbol{y}_h; \beta)^\top \boldsymbol{R}_h(\rho)^{-1/2} \boldsymbol{W}_h \boldsymbol{R}_h(\rho)^{-1/2} \boldsymbol{r}_h(\boldsymbol{y}_h; \beta)}{2s(x_h,\gamma)}\right],
\end{align*}
where $\boldsymbol{r}_h(\boldsymbol{y}_h; \beta)=\boldsymbol{y}_h-m(x_h,\beta)\boldsymbol{1}_{n_h}$. 
Then, the joint posterior distribution of the unknown parameters can be defined as 
\begin{equation*} \label{eq:Bayes_mPSUs}
\pi(\boldsymbol{\Theta})\prod_{h=1}^H L_w(\boldsymbol{y}_h; \beta, \gamma, \rho)
\prod_{l=1}^L \phi(\beta_{q+l}; 0, \tau_\beta^2)\phi(\gamma_{q+l}; 0, \tau_\gamma^2),
\end{equation*}
where $\boldsymbol{\Theta}=(\beta_0,\ldots,\beta_q, \gamma_0,\ldots,\gamma_q, \tau_\beta^2, \tau_\gamma^2, \rho)^\top$, and $\pi(\boldsymbol{\Theta})$ is its prior distribution. 

For parameters other than $\rho$, we use the same priors as in the previous section. We use a uniform prior distribution for $\rho$; that is, $\rho \sim \text{Uniform}(-1,1)$. 
We generate posterior samples via MCMC, where the details of the sampling steps are given in the supplementary material. 
Eventually, based on the posterior samples of $\gamma$, the overall variance can be calculated in the same way as in Section~\ref{sec:1PSU}.

Based on the posterior samples of $\gamma$ and $\rho$, the posterior samples of $s^2(x_h;\gamma) \boldsymbol{R}_h(\rho)$ can be obtained.
We use its posterior mean as its relevant point estimate.
The Bayesian variance estimator for the HT estimator can be obtained as
\begin{equation*} \label{eq:Bayesvar}
 var_{\text{Bayes}}= \frac{1}{N^2} \sum_{h=1}^{H} w^{\top}_{n_h} \hat{s}^2(x_h; \gamma) \hat{\boldsymbol{R}}_h(\rho) w_{n_h}, 
\end{equation*}
where $w^{\top}_{n_h} \propto (1/\pi_1 , ..., 1/\pi_{n_h})$ is the related vector of weights for multiple PSUs based on the sampling design.

%\textcolor{red}{
%\begin{myremark}
%    Note that for the case of multiple PSUs, $\pi_h$ in our Bayesian modeling framework associated with formula (\ref{eq:Bayes_mPSUs}) is the the probability of selecting multiple PSUs per stratum. For example, for the simple case of srswor related to our simulation studies in Section \ref{sec:normal} to follow, let us select $n_h=2$ PSUs per stratum with $N_h=60$, where $N_h$ is the size of the stratum at the population level. Then $\pi_h= n_h/N_h = 2/60 \approx 0.033$. 
%\end{myremark}
%}

\section{Simulation Studies} \label{sec:simulation}

In this section, we evaluate the performance of estimators using simulated data sets. In particular, we generate two data sets. For the first one, we assume that the super-population follows a normal distribution. For the second one, we assume that the super-population follows a gamma distribution. 

\subsection{Gaussian Super-Population} \label{sec:normal}

For the simulation study of this section, we follow the set up of \cite{breidt2016nonparametric} to generate a finite population. This set-up has previously been used by \cite{salinas2020calibration} to propose a new calibration estimator. It has also been used in a sequence of papers such as \cite{dahlke2013nonparametric}, \cite{diana2012calibration}, \cite{singh2012calibration}, and \cite{breidt2008endogenous}. We generate a finite stratified population with $H$ strata, each of them having the size of $N_h = NH^{-1} = 60$. For stratum $h$, we let $x_h = hH^{-1}$. We use the following mean function
\begin{equation*}
m_{*}(x)=2 \frac{g(x)-\min_{x \in [0,1]}g(x)}{\max_{x \in [0,1]}g(x)-\min_{x \in [0,1]}g(x)},
\end{equation*}  
where $g(x)=1+2(x-0.5)$.

In order to generate the survey variable, we let
$y_j=m_{*}(x_h)+e_j$ for $j=\{1,...,N_h\}$ such that $e_j \sim N(0,\phi^2)$. 
Additionally, we assume $\phi = \{0.25,0.5,5\}$, $H=\{50, 100,200\}$, and the sample size to be $n_h =1,2$ per stratum. 
We consider the HT estimator of the mean given in (\ref{eq:HTestimator}) and make comparisons among 3 different methods of estimating the variance as follows: 
1) collapsed variance estimator denoted by $var_{\text{Coll}}$, 2) nonparametric kernel-based variance estimator of \cite{breidt2016nonparametric} denoted by $var_{\text{Ker}}$, and 3) our proposed Bayesian variance estimator denoted by $var_{\text{Bayes}}$. 

The underlying sampling design is srswor, where all units have an equal chance of selection.
For collapsing the strata, we pair the adjacent strata based on the ordered values of $x_h$'s. Note that we did not collapse the strata for our proposed Bayesian estimator.
We conducted a simulation study with the size of $R=1000$ replications, where each replication contains an MCMC run of $10,000$ iterations with burn-in of $3000$. 

Our main evaluation criteria are as follows:
\begin{itemize}
    \item empirical absolute bias (AB) defined as $\text{AB}(var_{(.)})=|\mathbb{E}_{R}(var_{(.)}-V(\bar{y}_{\text{HT}}))|,$
    \item empirical root mean squared error (RMSE) defined as
    $\text{RMSE}(var_{(.)})=\{\mathbb{E}_{R}(var_{(.)}-V(\bar{y}_{\text{HT}}))^2\}^{1/2}$, and
    \item $95\%$ coverage probability (CP) defined as
     $\text{CP}(var_{(.)})= \frac{1}{R} \sum_{r=1}^{R} I_r$, where
     $$I_r = \begin{cases}
         1 & \text{if} \quad \bar{y} \in \text{CI} \\
         0 & \text{o.w.}
     \end{cases}$$ 
     and $\text{CI} = (\bar{y}_{\text{HT}} - 1.96 \sqrt{var_{(.)}},
     \bar{y}_{\text{HT}} + 1.96 \sqrt{var_{(.)}}).$

\end{itemize}

The results for the $H=50$ strata are given in Table \ref{Tab.Normal}. We observe that, as the value of $\phi$ and the number of PSUs increase, both the AB and the RMSE of the Bayesian variance estimator decrease more rapidly than those of the competing methods, leading to more accurate estimation of the true variance. In addition, the resulting CPs based on the Bayesian variance estimator are consistently closer to the nominal $95\%$ level. Additional simulation results are given in the supplementary material.

\begin{table}[ht]
\footnotesize
\caption{Comparisons among variance estimators from the Gaussian population. Note that $b=0.03$ for the kernel-based method and $L=7$ for the Bayesian method. We have $H=50$ strata.} \label{Tab.Normal}
\centering
\setlength{\tabcolsep}{4pt} % Default value: 6pt
\renewcommand{\arraystretch}{1.45} % Default value: 1
\begin{tabular}{@{} cccccc @{}}
\hline
$\phi$ & PSU & Estimator & AB & RMSE & $95\%$ CP  \\ \hline\hline
0.25 & 1 & $var_{\text{Coll}}$ & 3e-04 & 4e-04 & 0.931 \\
 && $var_{\text{Ker}}$ & 3e-04 & 4e-04 & 0.930 \\
 && $var_{\text{Bayes}}$ & 3e-04 & 4e-04 & 0.957 \\ \hline
 & 2 & $var_{\text{Coll}}$ & 7.417e-05 & 9.304e-05 & 0.945 \\
 && $var_{\text{Ker}}$ & 1.300e-04 & 1.684e-04 & 0.943 \\
 && $var_{\text{Bayes}}$ & 7.159e-05 & 9.033e-05 & 0.956 \\
\hline \hline
0.5 & 1 & $var_{\text{Coll}}$ & 0.001 & 0.001 & 0.932 \\
 && $var_{\text{Ker}}$ & 0.001 & 0.001 & 0.933 \\
 && $var_{\text{Bayes}}$ & 0.001 & 0.001 & 0.952 \\ \hline
 & 2 & $var_{\text{Coll}}$ & 3e-04 & 4e-04 & 0.944 \\
 && $var_{\text{Ker}}$ & 5e-04 & 7e-04 & 0.943 \\
 && $var_{\text{Bayes}}$ & 3e-04 & 3e-04 & 0.956 \\
\hline \hline
5 & 1 & $var_{\text{Coll}}$ & 0.111 & 0.139 & 0.928 \\
 && $var_{\text{Ker}}$ & 0.110 & 0.140 & 0.930 \\
 && $var_{\text{Bayes}}$ & 0.077 & 0.095 & 0.930 \\ \hline
 & 2 & $var_{\text{Coll}}$ & 0.030 & 0.037 & 0.947 \\
 && $var_{\text{Ker}}$ & 0.055 & 0.070 & 0.939 \\
 && $var_{\text{Bayes}}$ & 0.026 & 0.032 & 0.945 \\
\hline
\end{tabular}
\end{table}

\subsection{HMT Population} \label{sec:HMT}

In order to conduct the simulation study for a super-population different from the Gaussian, we use $\texttt{HMT\,\{PracTools}\}$ in \textsf{R} to generate a population based on the gamma distributions for $(x_h,y_h)$ where the values of $x_h$ differ per unit (see \cite{valliant2020package} for further details). Using this function, strata are formed to approximately have the same total of $x$. The \textit{HMT population} was first introduced by \cite{hansen1983evaluation}. 
Here, we generate $N=2000$ observations with $H=20$ strata. Then, we assume $x_h=\{x_h-\min(x) \} / \{\max(x)- \min(x) \}$.

We have considered two underlying sampling designs: srswor and systematic, where the inclusion probability differs per unit under the systematic case. %(see the available code for the further details). 
We consider the same evaluation criteria used in Section \ref{sec:normal}. The results are given in Tables \ref{Tab.HMTsrswor} and \ref{Tab.HMTsystematic}. Consistent with the results in the previous section, we find that the Bayesian variance estimator yields more accurate variance estimates -characterized by smaller AB and RMSE- particularly under systematic sampling designs, when compared with the collapsed variance estimator. Moreover, the CPs associated with the Bayesian estimator are more stable across scenarios and remain closer to the nominal $95\%$ level than those obtained using alternative methods.

\begin{table}[ht]
\footnotesize
\caption{Comparisons among variance estimators from the HMT population. The underlying design is srswor. Note that the bandwidth is $b=0.06$ for the kernel-based method and $L=7$ for the Bayesian method.} \label{Tab.HMTsrswor}
\centering
\setlength{\tabcolsep}{6pt} % Default value: 6pt
\renewcommand{\arraystretch}{1.15} % Default value: 1
\begin{tabular}{@{} ccccc @{}}
\hline
PSU & Estimator & AB & RMSE & $95\%$ CP  \\ \hline\hline
1 & $var_{\text{Coll}}$ & 0.044  & 0.061 & 0.927 \\
 & $var_{\text{Ker}}$ & 0.041 & 0.054  & 0.932 \\
 & $var_{\text{Bayes}}$ & 0.030  & 0.040 & 0.912 \\
 \hline
2 & $var_{\text{Coll}}$ & 0.014  & 0.019 & 0.937 \\
 & $var_{\text{Ker}}$ & 0.023 & 0.032 & 0.911 \\
 & $var_{\text{Bayes}}$ & 0.011 & 0.015 & 0.934 \\
\hline
\end{tabular}
\end{table}

\begin{table}[ht]
\footnotesize
\caption{Comparisons among variance estimators from the HMT population. The underlying design is systematic. Note that the bandwidth is $b=0.06$ for the kernel-based method and $L=7$ for the Bayesian method.} \label{Tab.HMTsystematic}
\centering
\setlength{\tabcolsep}{6pt} % Default value: 6pt
\renewcommand{\arraystretch}{1.15} % Default value: 1
\begin{tabular}{@{} ccccc @{}}
\hline
PSU & Estimator & AB & RMSE & $95\%$ CP  \\ \hline\hline
1 & $var_{\text{Coll}}$ & 0.164 & 0.175 & 0.997 \\
 & $var_{\text{Ker}}$ & 0.040 & 0.051 & 0.925 \\
 & $var_{\text{Bayes}}$ & 0.035 & 0.043 & 0.921 \\
 \hline
2 & $var_{\text{Coll}}$ & 0.190 & 0.193 & 0.999 \\
 & $var_{\text{Ker}}$ & 0.024 & 0.030 & 0.905 \\
 & $var_{\text{Bayes}}$ & 0.015  & 0.024 & 0.968 \\
\hline
\end{tabular}
\end{table}

\section{Illustrative Analysis Using Data from the National Survey of Family Growth} \label{sec:application}

The National Survey of Family Growth (NSFG) is a survey on fertility, family formation and change, family planning, reproductive health, and closely related topics. It is a major source of national estimates on a variety of fertility and family topics. The target population for the NSFG consists of all non-institutionalized women and men aged 15-49 years as of first contact for the survey, living in households, and whose usual place of residence is the 50 United States and the District of Columbia.

The NSFG is designed and administered by the National Center for Health Statistics (NCHS) (see \href{https://www.cdc.gov/nchs/nsfg/about_nsfg.htm}{https://www.cdc.gov/nchs/nsfg/about-nsfg.htm} for more information). The sample design for the 2015-2017 NSFG is based on a stratified multi-stage area probability sample of the household population aged 15-49. The sample design is quite complex, and an interested reader can review the related documentation (\href{https://www.cdc.gov/nchs/data/nsfg/PUF3-NSFG-2015-2017-Sample-Design-Documentation_26Sept2019.pdf}{Documents}).

The public-use data set that we use here is related to the two-year (2015--2017) file. We used design variables for the sampling stratum (SEST) and the cluster (SECU) to obtain the correct standard errors for the estimates. There are 4 clusters per stratum and there are 18 strata. We assume each cluster to be one PSU.
The sample design and fieldwork for the NSFG were conducted by the University of Michigan Institute for Social Research.
The female pregnancy file has 9,553 observations and 248 variables. Based on reported data, the total number of PSUs is N=2,149.

We used 3 variables related to the pregnancy of the women interviewed in our study: (1) ``AGEPREG" (pregnancy age), (2) ``EDUCAT" (number of years of schooling) and (3) ``NBRNALIV" (number of babies). In addition, we used ``POVERTY" (poverty-level income) for $x$, and standardized it across all strata.
In order to describe our methodology, we consider two scenarios:
\begin{itemize}
    \item 1 PSU: We have picked one PSU per stratum with the largest size.
    \item 2 PSUs: We have picked two PSUs per stratum with the largest sizes.
\end{itemize}

\begin{table}[ht]
\footnotesize
\caption{Comparisons among variance estimators for the NSFG} \label{Tab.NSFG}
\centering
\setlength{\tabcolsep}{6pt} % Default value: 6pt
\renewcommand{\arraystretch}{1.15} % Default value: 1
\begin{tabular}{@{} ccccc @{}}
\hline
PSU & Estimator & Pregnancy Age & Years of Schooling & Number of Babies  \\ \hline\hline
1 & $var_{\text{Coll}}$ & 1.592e-01 & 4.499e-02 & 2.690e-04 \\
 & $var_{\text{Ker}}$ & 1.906e-02 & 4.835e-03 & 3.291e-05 \\
 & $var_{\text{Bayes}}$ & 3.412e-04 & 1.540e-04 & 7.575e-08 \\
 \hline
2 & $var_{\text{Coll}}$ & 7.507e-01 & 2.041e-01 & 1.250e-03 \\
 & $var_{\text{Ker}}$ & 1.786e-02 & 2.057e-03 & 1.754e-05 \\
 & $var_{\text{Bayes}}$ & 1.299e-03 & 4.242e-04 & 3.353e-07 \\
\hline
\end{tabular}
\end{table}

Then, we assume the sufficient statistic corresponding to average for $y_h$ and $x_h$ per PSU and calculate the HT estimator based on equation (\ref{eq:HTestimator}) for the 3 variables as well as their variances based on the 3 methods described in this manuscript.  The numerical values of the variances are given in Table \ref{Tab.NSFG}. 

\begin{figure}[ht]
\centering
\begin{tabular}{ cc }
\includegraphics[width=0.5\textwidth]{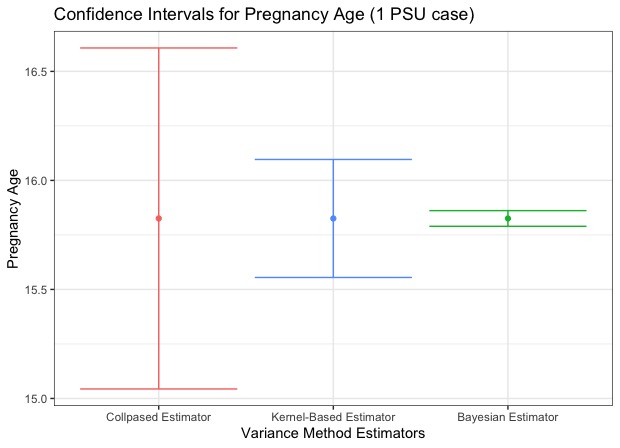} &
\includegraphics[width=0.5\textwidth]{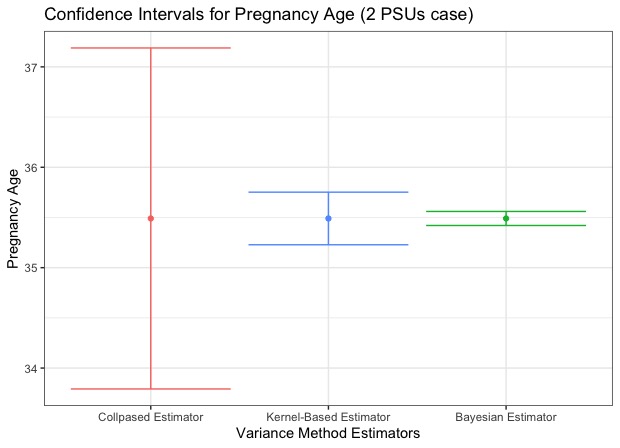} \\
\includegraphics[width=0.5\textwidth]{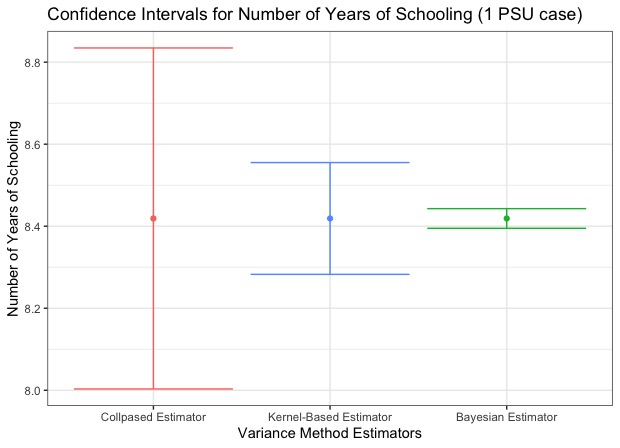} &
\includegraphics[width=0.5\textwidth]{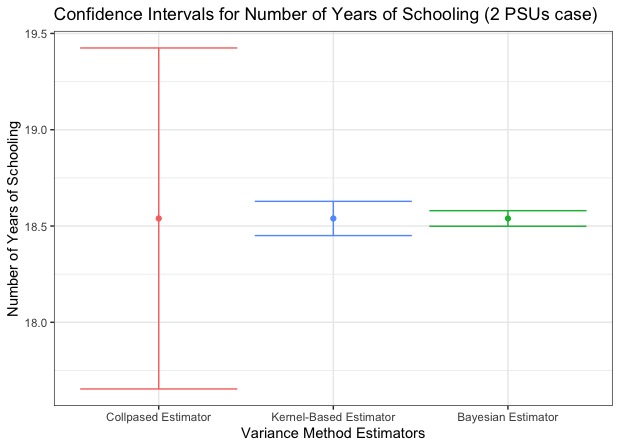} \\
\includegraphics[width=0.5\textwidth]{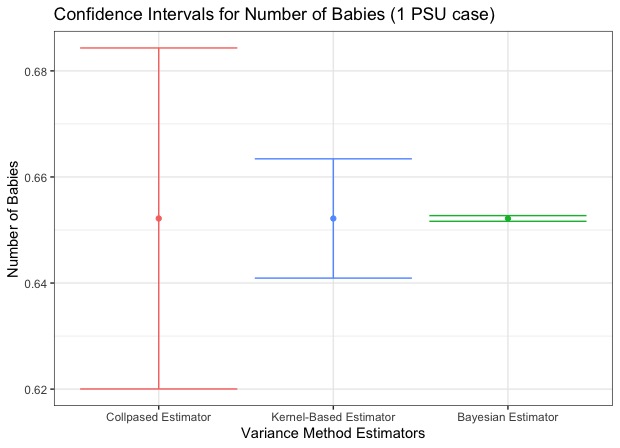} &
\includegraphics[width=0.5\textwidth]{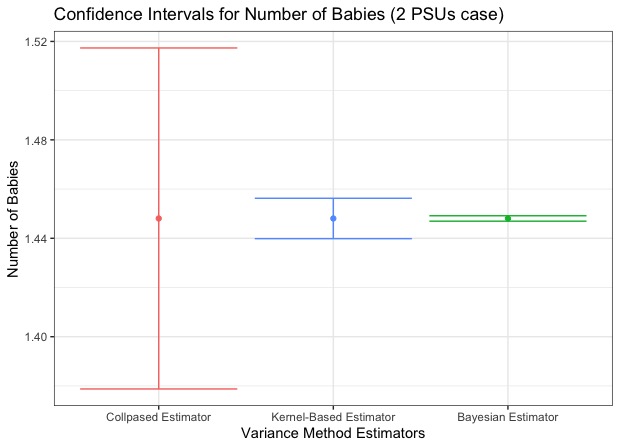}
\end{tabular}
\caption{Confidence intervals of multiple variables from NSFG. Note that for the sake of figures readability, the HT estimators are rescaled by a factor of 7.}
\label{CIs_NSFG}
\end{figure}

We observe that the estimated variances based on our proposed Bayesian method are relatively smaller than the rest. Furthermore, we calculate the $95\%$ confidence intervals using all variances, and the results are displayed in Figure \ref{CIs_NSFG}. The resulting intervals based on our proposed method are shorter than those obtained from alternative approaches. We emphasize that the survey weights in this data set are variable and that the total number of PSUs is quite large, with final inclusion probabilities ranging from 
$0.04$ to nearly $0.20$, which can lead to increased posterior concentration. We view this behavior as a general characteristic of weighted likelihood-based approaches, rather than as a fundamental limitation of the proposed method, and as a consequence of jointly minimizing mean and variance.

\section{Discussion} \label{sec:discuss}

Fine stratification is a popular design, as it allows the stratification to be carried out to the fullest possible extent.
In this article, we propose a Bayesian variance estimator that relies on the penalized spline method for the fine stratification design. This method allows us to smooth the means and variances simultaneously across all strata. Unlike the alternative methods described in the paper, our method does not rely on collapsing strata and can be easily used by government agencies.
This modeling framework is flexible and can be extended to the cases where more than one auxiliary variable exists per stratum, which is left to future work.

Often, in the applications, the real PSU is a group of units such as a hospital or a school. Thus, $y_h$ can serve as a sufficient statistic for each PSU, similar to the NSFG example, and we can assume that their collections, as stated in Equations (\ref{eq:model-PSU1}) and (\ref{eq:N_multiple}) follow the normal assumption.
Small to moderate deviations (mild skewness or heavier-than-normal tails) typically have little effect on standard inferential procedures, especially as the number of strata increases due to the CLT. However, the assumption is not robust to strong deviations such as extreme skewness, multimodality, or heavy tails, particularly when the number of strata is small. For these cases, the assumption of normality could be changed, although our main methodology remains the same. 

Although the use of $\boldsymbol{R}_h(\rho)$ to incorporate correlations among multiple PSUs can be useful, the assumption of equicorrelation, as considered in Section~\ref{sec:multi-PSU} is mainly for mathematical convenience, which may provide some benefit even when the correlation structure is misspecified similarly to the working correlation in the generalized estimating equation \citep{liang1986longitudinal}. 
When some prior knowledge on the correlation structure is available, we can use other correlation matrices such as autoregressive correlation. 

While the proposed Bayesian framework is primarily used in this paper to obtain point estimates of variance parameters, the full posterior distribution contains additional information that could be exploited in future work. In particular, posterior uncertainty in the variance estimates could be used to characterize the variability of the variance estimator itself, which is closely related to the concept of effective degrees of freedom commonly used in survey sampling and variance estimation. Under classical approximations, variance estimators are often linked to chi-square distributions, with the associated degrees of freedom determining the accuracy of confidence intervals and hypothesis tests. An interesting extension of the proposed method would be to derive an approximate degrees-of-freedom measure based on posterior variability and to compare it with existing design-based or collapsed-strata approaches. Such developments could further enhance the usefulness of the proposed method for downstream inference, including confidence interval construction and formal statistical testing.

Finally, the proposed methodology focuses on univariate variance estimation, where modeling the log-variance function ensures non-negativity in a straightforward manner. An important direction for future work is the extension of this framework to multivariate settings, such as joint variance estimation for multiple HT estimators. In such cases, direct spline-based modeling of covariance components may not automatically guarantee positive definiteness of the resulting covariance matrix, even when all marginal variances are positive. Addressing this issue would require additional structural constraints. Possible approaches include modeling the covariance matrix through structured parameterizations, such as autoregressive or random-walk priors, or adopting transformations that enforce positive definiteness by construction. Alternatively, kernel-based or weighted averaging approaches to covariance estimation preserve positive definiteness when combining positive definite stratum-level contributions using non-negative weights. Exploring and comparing these strategies within a Bayesian framework represents a promising avenue for future research.

\clearpage

\bibliographystyle{ims}
\bibliography{Bibliography}

\begin{thebibliography}{38}
\expandafter\ifx\csname natexlab\endcsname\relax\def\natexlab#1{#1}\fi
\expandafter\ifx\csname url\endcsname\relax
  \def\url#1{\texttt{#1}}\fi
\expandafter\ifx\csname urlprefix\endcsname\relax\def\urlprefix{URL }\fi
\providecommand{\eprint}[2][]{\url{#2}}

\bibitem[{Bissiri et~al.(2016)Bissiri, Holmes and Walker}]{bissiri2016general}
{Bissiri, P.~G.}, {Holmes, C.~C.} and {Walker, S.~G.} (2016)
\newblock ``{A} general framework for updating belief distributions,"
\newblock \textit{Journal of the Royal Statistical Society Series B:
  Statistical Methodology,} \text{78}, 1103--1130.

\bibitem[{Bouezmarni and Scaillet(2005)}]{bouezmarni2005consistency}
{Bouezmarni, T.} and {Scaillet, O.} (2005)
\newblock ``{C}onsistency of asymmetric kernel density estimators and smoothed
  histograms with application to income data,"
\newblock \textit{Econometric Theory,} \text{21}, 390--412.

\bibitem[{Breidt(2002)}]{Breidt2002}
{Breidt, F.~J.} (2002)
\newblock ``{T}he national resources inventory,"
\newblock \textit{Encyclopedia of Environmetrics,} 135--136.

\bibitem[{Breidt and Opsomer(2008)}]{breidt2008endogenous}
{Breidt, F.~J.} and {Opsomer, J.~D.} (2008)
\newblock ``{E}ndogenous post-stratification in surveys: classifying with a
  sample-fitted model,"
\newblock \textit{The Annals of Statistics,} \text{36}, 403--427.

\bibitem[{Breidt et~al.(2016)Breidt, Opsomer and
  Sanchez-Borrego}]{breidt2016nonparametric}
{Breidt, F.~J.}, {Opsomer, J.~D.} and {Sanchez-Borrego, I.} (2016)
\newblock ``{N}onparametric variance estimation under fine stratification: An
  alternative to collapsed strata,"
\newblock \textit{Journal of the American Statistical Association,} \text{111},
  822--833.

\bibitem[{{Census Bureau}(2006)}]{USCB2006}
{{Census Bureau}} (2006)
\newblock \textit{Current Population Survey Design and Methodology}
\newblock Technical Paper 66, Washington, DC.

\bibitem[{Dahlke et~al.(2013)Dahlke, Breidt, Opsomer and
  Van~Keilegom}]{dahlke2013nonparametric}
{Dahlke, M.}, {Breidt, F.~J.}, {Opsomer, J.~D.} and {Van~Keilegom, I.} (2013)
\newblock ``{N}onparametric endogenous post-stratification estimation,"
\newblock \textit{Statistica Sinica,} \text{23}, 189--211.

\bibitem[{Diana and Perri(2012)}]{diana2012calibration}
{Diana, G.} and {Perri, P.~F.} (2012)
\newblock ``{A} calibration-based approach to sensitive data: a simulation
  study,"
\newblock \textit{Journal of Applied Statistics,} \text{39}, 53--65.

\bibitem[{Godambe and Thompson(1986)}]{godambe1986parameters}
{Godambe, V.} and {Thompson, M.~E.} (1986)
\newblock ``{P}arameters of superpopulation and survey population: Their
  relationships and estimation,"
\newblock \textit{International Statistical Review,} 127--138.

\bibitem[{Hansen et~al.(1953)Hansen, Hurwitz and Madow}]{hansen1953sample}
{Hansen, M.~H.}, {Hurwitz, W.~N.} and {Madow, W.~G.} (1953)
\newblock \textit{Sample survey methods and theory Volume I}
\newblock John Wiley.

\bibitem[{Hansen et~al.(1983)Hansen, Madow and Tepping}]{hansen1983evaluation}
{Hansen, M.~H.}, {Madow, W.~G.} and {Tepping, B.~J.} (1983)
\newblock ``{A}n evaluation of model-dependent and probability-sampling
  inferences in sample surveys,"
\newblock \textit{Journal of the American Statistical Association,} \text{78},
  776--793.

\bibitem[{Horvitz and Thompson(1952)}]{horvitz1952generalization}
{Horvitz, D.~G.} and {Thompson, D.~J.} (1952)
\newblock ``{A} generalization of sampling without replacement from a finite
  universe,"
\newblock \textit{Journal of the American statistical Association,} \text{47},
  663--685.

\bibitem[{Isaki(1983)}]{isaki1983variance}
{Isaki, C.~T.} (1983)
\newblock ``{V}ariance estimation using auxiliary information,"
\newblock \textit{Journal of the American Statistical Association,} \text{78},
  117--123.

\bibitem[{Lepkowski et~al.(2010)Lepkowski, Mosher, Davis, Groves and
  Van~Hoewyk}]{lepkowski20102006}
{Lepkowski, J.~M.}, {Mosher, W.~D.}, {Davis, K.~E.}, {Groves, R.~M.} and
  {Van~Hoewyk, J.} (2010)
\newblock ``{T}he 2006-2010 national survey of family growth: sample design and
  analysis of a continuous survey,"
\newblock \textit{Vital and health statistics. Series 2, Data evaluation and
  methods research,} 1--36.

\bibitem[{Liang and Zeger(1986)}]{liang1986longitudinal}
{Liang, K.-Y.} and {Zeger, S.~L.} (1986)
\newblock ``{L}ongitudinal data analysis using generalized linear models,"
\newblock \textit{Biometrika,} \text{73}, 13--22.

\bibitem[{Lohr(1999)}]{Lohr1999}
{Lohr, S.~L.} (1999)
\newblock \textit{Sampling: Design and Analysis}
\newblock Pacific Grove, CA: Duxbury Press.

\bibitem[{Mantel and Giroux(2009)}]{harold2009variance}
{Mantel, H.} and {Giroux, S.} (2009)
\newblock ``{V}ariance estimation in complex surveys with one psu per stratum,"
\newblock \textit{Washington, DC: JSM Proceedings, Statistical Computing
  Section}.

\bibitem[{Mosaferi(2015)}]{mosaferiempirical}
{Mosaferi, S.} (2015)
\newblock ``{E}mpirical and constrained empirical {B}ayes variance estimation
  under a one unit per stratum sample design,"
\newblock \textit{American Statistical Association Proceedings of the Survey
  Research Methods Section,} 2452--2464.

\bibitem[{Nott(2006)}]{nott2006semiparametric}
{Nott, D.} (2006)
\newblock ``{S}emiparametric estimation of mean and variance functions for
  non-gaussian data,"
\newblock \textit{Computational Statistics,} \text{21}, 603--620.

\bibitem[{Nusser and Goebel(1997)}]{nusser1997national}
{Nusser, S.~M.} and {Goebel, J.~J.} (1997)
\newblock ``{T}he national resources inventory: a long-term multi-resource
  monitoring programme,"
\newblock \textit{Environmental and Ecological Statistics,} \text{4}, 181--204.

\bibitem[{Opsomer et~al.(2008)Opsomer, Claeskens, Ranalli, Kauermann and
  Breidt}]{opsomer2008non}
{Opsomer, J.~D.}, {Claeskens, G.}, {Ranalli, M.~G.}, {Kauermann, G.} and
  {Breidt, F.~J.} (2008)
\newblock ``{N}on-parametric small area estimation using penalized spline
  regression,"
\newblock \textit{Journal of the Royal Statistical Society Series B:
  Statistical Methodology,} \text{70}, 265--286.

\bibitem[{Pati and Dunson(2014)}]{pati2014bayesian}
{Pati, D.} and {Dunson, D.~B.} (2014)
\newblock ``{B}ayesian nonparametric regression with varying residual density,"
\newblock \textit{Annals of the Institute of Statistical Mathematics,}
  \text{66}, 1--31.

\bibitem[{Pfeffermann(1993)}]{pfeffermann1993role}
{Pfeffermann, D.} (1993)
\newblock ``{T}he role of sampling weights when modeling survey data,"
\newblock \textit{International Statistical Review,} 317--337.

\bibitem[{Ruppert(2002)}]{ruppert2002selecting}
{Ruppert, D.} (2002)
\newblock ``{S}electing the number of knots for penalized splines,"
\newblock \textit{Journal of Computational and Graphical Statistics,}
  \text{11}, 735--757.

\bibitem[{Rust and Kalton(1987)}]{rust1987strategies}
{Rust, K.} and {Kalton, G.} (1987)
\newblock ``{S}trategies for collapsing strata for variance estimation,"
\newblock \textit{Journal of Official Statistics,} \text{3}, 69--81.

\bibitem[{Salinas et~al.(2020)Salinas, Sedory and
  Singh}]{salinas2020calibration}
{Salinas, V.~I.}, {Sedory, S.~A.} and {Singh, S.} (2020)
\newblock ``{C}alibration using power transformation,"
\newblock \textit{Communications in Statistics-Simulation and Computation,}
  \text{49}, 2256--2286.

\bibitem[{S{\"a}rndal et~al.(2003)S{\"a}rndal, Swensson and
  Wretman}]{sarndal2003model}
{S{\"a}rndal, C.~E.}, {Swensson, B.} and {Wretman, J.} (2003)
\newblock \textit{Model assisted survey sampling}
\newblock Springer Science \& Business Media.

\bibitem[{Savitsky and Toth(2016)}]{savitsky2016bayesian}
{Savitsky, T.~D.} and {Toth, D.} (2016)
\newblock ``{B}ayesian estimation under informative sampling,"
\newblock \textit{Electronic Journal of Statistics,} \text{10}, 1677--1708.

\bibitem[{Singh(2012)}]{singh2012calibration}
{Singh, S.} (2012)
\newblock ``{O}n the calibration of design weights using a displacement
  function,"
\newblock \textit{Metrika,} \text{75}, 85--107.

\bibitem[{Sugasawa and Kim(2022)}]{sugasawa2022approximate}
{Sugasawa, S.} and {Kim, J.~K.} (2022)
\newblock ``{A}n approximate {B}ayesian approach to model-assisted survey
  estimation with many auxiliary variables,"
\newblock \textit{Statistica Sinica,} \text{32}, 477--498.

\bibitem[{Valliant et~al.(2013)Valliant, Dever and
  Kreuter}]{valliant2013practical}
{Valliant, R.}, {Dever, J.~A.} and {Kreuter, F.} (2013)
\newblock \textit{Practical tools for designing and weighting survey samples}
\newblock Springer.

\bibitem[{Valliant et~al.(2020)Valliant, Dever and
  Kreuter}]{valliant2020package}
{Valliant, R.}, {Dever, J.~A.} and {Kreuter, F.} (2020)
\newblock Package ‘practools’.

\bibitem[{Wang et~al.(2018)Wang, Kim and Yang}]{wang2018approximate}
{Wang, Z.}, {Kim, J.} and {Yang, S.} (2018)
\newblock ``{A}pproximate {B}ayesian inference under informative sampling,"
\newblock \textit{Biometrika,} \text{105}, 91--102.

\bibitem[{Westat(2001)}]{Westat2001}
{Westat} (2001)
\newblock \textit{Survey of Income and Program Participation Users' Guide}
\newblock (3rd ed.), Technical Report, Rockville, MD.

\bibitem[{Williams and Savitsky(2021)}]{williams2021uncertainty}
{Williams, M.~R.} and {Savitsky, T.~D.} (2021)
\newblock ``{U}ncertainty estimation for pseudo-{B}ayesian inference under
  complex sampling,"
\newblock \textit{International Statistical Review,} \text{89}, 72--107.

\bibitem[{Wolter(2007)}]{wolter2007introduction}
{Wolter, K.} (2007)
\newblock \textit{Introduction to variance estimation}
\newblock Springer Science \& Business Media.

\bibitem[{Zhao et~al.(2020)Zhao, Ghosh, Rao and Wu}]{zhao2020bayesian}
{Zhao, P.}, {Ghosh, M.}, {Rao, J.} and {Wu, C.} (2020)
\newblock ``{B}ayesian empirical likelihood inference with complex survey
  data,"
\newblock \textit{Journal of the Royal Statistical Society Series B:
  Statistical Methodology,} \text{82}, 155--174.

\bibitem[{Zheng and Little(2003)}]{zheng2003penalized}
{Zheng, H.} and {Little, R.~J.} (2003)
\newblock ``{P}enalized spline model-based estimation of the finite populations
  total from probability-proportional-to-size samples,"
\newblock \textit{Journal of official Statistics} \text{19}, 99--117.

\end{thebibliography}

%%%%%%%%%% Merge with supplemental materials %%%%%%%%%%
\newpage
\clearpage

\setcounter{page}{1}
\setcounter{section}{0}
\setcounter{table}{0}
\renewcommand{\thesection}{\Alph{section}}

\begin{center}
\LARGE{Supplementary Material for \hspace{.2cm}\\ ``Bayesian Estimation of Variance under Fine Stratification via Mean-Variance Smoothing"}
\vspace{1cm}
\end{center}

\noindent This supplementary material is structured as follows. In Section A, we provide some detailed derivations related to the conditional distributions. In Section B, we provide additional simulation results. In Section C, we give details of pseudo likelihood and Section D explains its effect on the results through some simulation studies.

\section{Details of Full Conditional Distributions for MCMC} \label{App_A}

\subsection{Single PSU per Stratum}
Note that the joint posterior is expressed as  
\begin{equation*}
\pi(\boldsymbol{\Theta})\prod_{h=1}^H \phi(y_h; m(x_h; \beta), s^2(x_h; \gamma))^{\tilde{w}_h}\prod_{l=1}^L \phi(\beta_{q+l}; 0, \tau_\beta^2)\phi(\gamma_{q+l}; 0, \tau_\gamma^2).
\end{equation*}
The detailed sampling steps are given as follows: 

\begin{itemize}
\item[-]
(Sampling from $\tau_{\beta}^2$) \ \ 
The full conditional distribution of $\tau_{\beta}^2$ is ${\rm IG}(a_{\beta}+L/2, b_{\beta}+\sum_{l=1}^L\beta_{q+l}^2/2)$. 

\item[-]
(Sampling from $\tau_{\gamma}^2$) \ \ 
The full conditional distribution of $\tau_{\gamma}^2$ is ${\rm IG}(a_{\gamma}+L/2, b_{\gamma}+\sum_{l=1}^L\gamma_{q+l}^2/2)$. 

\item[-]
(Sampling from $\beta$) \ \ 
The full conditional distribution of $\beta$ is $N(A_{\beta}^{-1}B_{\beta}, A_{\beta}^{-1})$, where 

$$
A_{\beta}={\rm diag}(S_\beta^{-1}I_{q+1}, \tau_{\beta}^{-2}I_L)+\sum_{h=1}^H \frac{\tilde{w}_h z_{h}z_h^\top}{s^2(x_h; \gamma)}, \ \ \ \ \ 
B_{\beta}=\sum_{h=1}^H \frac{\tilde{w}_h z_{h}y_h}{s^2(x_h; \gamma)}
$$
with $z_h=(1,x_h,\ldots,x_h^q, (x-\kappa_1)_{+}^q, \ldots,(x-\kappa_L)_{+}^q)$.

\item[-]
(Sampling from $\gamma$) \ \ 
The log-density of the full conditional distribution of $\gamma$ without irrelevant constant is 
$$
\log p_f(\gamma)\equiv \sum_{h=1}^H \tilde{w}_h \log \phi(y_h; m(x_h; \beta), s^2(x_h; \gamma)) - \frac12\sum_{k=0}^q \frac{\gamma_k^2}{S_\gamma} - \frac12\sum_{l=1}^L\frac{\gamma_{q+l}^2}{\tau_\gamma^2},
$$
for which we employ a random-walk Metropolis-Hasting algorithm. 
Given the current value, say $\gamma^{\dagger}$, we generate a proposal $\gamma^{\ast}$ from $N(\gamma^{\dagger}, bI_{q+L})$ with fixed variance $b$ and accept the proposal with probability ${\rm min}\left\{1, p_f(\gamma^{\ast})/p_f(\gamma^{\dagger})\right\}$.
\end{itemize}

\subsection{Multiple PSUs per Stratum}
Note that the joint posterior is expressed as 
\begin{equation*}
\pi(\boldsymbol{\Theta})\prod_{h=1}^H L_w(\boldsymbol{y}_h; \beta, \gamma, \rho)
\prod_{l=1}^L \phi(\beta_{q+l}; 0, \tau_\beta^2)\phi(\gamma_{q+l}; 0, \tau_\gamma^2),
\end{equation*}
where 
\begin{align*}
L_w(\boldsymbol{y}_h; \beta, \gamma, \rho)
&=|\boldsymbol{R}_h(\rho)|^{-1/2}s(x_h,\gamma)^{ -\sum_{j=1}^{n_h}\tilde{w}_{hj}/n_h} \exp\left[-\frac{\boldsymbol{r}_h(\boldsymbol{y}_h; \beta)^\top \boldsymbol{R}_h(\rho)^{-1/2} \boldsymbol{W}_h \boldsymbol{R}_h(\rho)^{-1/2} \boldsymbol{r}_h(\boldsymbol{y}_h; \beta)}{2s(x_h,\gamma)}\right],
\end{align*}
with $\boldsymbol{r}_h(\boldsymbol{y}_h; \beta)=\boldsymbol{y}_h-m(x_h,\beta)\boldsymbol{1}_{n_h}$. 
The detailed sampling steps are given as follows: 

\begin{itemize}
\item[-]
(Sampling from $\tau_{\beta}^2$ and $\tau^2_{\gamma}$) \ \ 
The full conditional distributions of $\tau^2_{\beta}$ and $\tau^2_{\gamma}$ are exactly the same as those for the case with a single PSU.

\item[-]
(Sampling from $\beta$) \ \ 
The full conditional distribution of $\beta$ is 
$N(A_{\beta}^{-1}B_{\beta}, A_{\beta}^{-1})$, where 
\begin{align*}
A_{\beta}&={\rm diag}(S_\beta^{-1}I_{q+1}, \tau_{\beta}^{-2}I_L)+\sum_{h=1}^H \frac{\boldsymbol{z}_{h} \boldsymbol{1}_{n_h}^{\top} \boldsymbol{R}_h(\rho)^{-1/2} \boldsymbol{W}_h \boldsymbol{R}_h(\rho)^{-1/2}  \boldsymbol{1}_{n_h}\boldsymbol{z}_h^\top}{s^2(x_h; \gamma)}, \\
B_{\beta}&=\sum_{h=1}^H \frac{\boldsymbol{z}_{h} \boldsymbol{1}_{n_h}^{\top}
\boldsymbol{R}_h(\rho)^{-1/2} \boldsymbol{W}_h \boldsymbol{R}_h(\rho)^{-1/2} \boldsymbol{1}_{n_h}\boldsymbol{y}_h}{s^2(x_h; \gamma)}.
\end{align*}

\item[-]
(Sampling from $\gamma$) \ \ 
The log-density of the full conditional distribution of $\gamma$ without irrelevant constant is 
\begin{align*}
\log p_f(\gamma)
&\equiv 
-\frac12\sum_{h=1}^H\frac{\boldsymbol{r}_h(\boldsymbol{y}_h; \beta)^\top \boldsymbol{R}_h(\rho)^{-1/2} \boldsymbol{W}_h \boldsymbol{R}_h(\rho)^{-1/2} \boldsymbol{r}_h(\boldsymbol{y}_h; \beta)}{2s(x_h,\gamma)}\\
& \ \ \ 
-\frac12\sum_{h=1}^H  \sum_{j=1}^{n_h} \tilde{w}_{hj}\log s(x_h, \gamma)
- \frac12\sum_{k=0}^q \frac{\gamma_k^2}{S_\gamma} - \frac12\sum_{l=1}^L\frac{\gamma_{q+l}^2}{\tau_\gamma^2},
\end{align*}
for which we employ a random-walk Metropolis-Hasting algorithm. 
Given the current value, say $\gamma^{\dagger}$, we generate a proposal $\gamma^{\ast}$ from $N(\gamma^{\dagger}, b_\gamma I_{q+L})$ with fixed variance $b_\gamma$ and accept the proposal with probability ${\rm min}\left\{1, p_f(\gamma^{\ast})/p_f(\gamma^{\dagger})\right\}$.

\item[-]
(Sampling from $\rho$) \ \ 
The log-density of the full conditional distribution of $\rho$ without irrelevant constant is 
$$
\log p_f(\rho)\equiv
-\frac12\sum_{h=1}^H\frac{\boldsymbol{r}_h(\boldsymbol{y}_h; \beta)^\top \boldsymbol{R}_h(\rho)^{-1/2} \boldsymbol{W}_h \boldsymbol{R}_h(\rho)^{-1/2} \boldsymbol{r}_h(\boldsymbol{y}_h; \beta)}{2s(x_h,\gamma)} 
-\frac{1}{2}\sum_{h=1}^H \log |\boldsymbol{R}_h(\rho)|,
$$
for which we again use a random-walk Metropolis-Hasting algorithm. 
Given the current value, say $\rho^{\dagger}$, we generate a proposal $\rho^{\ast}$ from $N(\gamma^{\dagger}, b_\rho)$ with fixed variance $b_\rho$ and accept the proposal with probability ${\rm min}\left\{1, p_f(\rho^{\ast})/p_f(\rho^{\dagger})\right\}$.
\end{itemize}

\clearpage

\section{Additional Simulation Results} \label{App_B}
\renewcommand{\theequation}{B.\arabic{equation}}
\setcounter{equation}{0}
\renewcommand{\thelemma}{B.\arabic{lemma}}
\renewcommand{\thedefinition}{B.\arabic{definition}}
\renewcommand{\thetable}{B.\arabic{table}}
\setcounter{figure}{0}

In this section, we provide additional simulation results for the Gaussian super-population. In Tables \ref{Tab.B1} and \ref{Tab.B2}, we give simulation results for $H=100$ and $H=200$ strata, respectively. The results show that the $var_{\text{Bayes}}$ estimator can perform significantly better than the alternative estimators. Furthermore, in Table \ref{Tab.B3}, we investigate the performance of our variance estimator $var_{\text{Bayes}}$ on different values of $L=(5,7,10)$. We observe that the results do not differ significantly between the values of $L$.    

\begin{table}[ht]
\footnotesize
\caption{Comparisons among variance estimators from the Gaussian population. Note that $b=0.015$ for the kernel-based method and $L=7$ for the Bayesian method. We have $H=100$ strata.} \label{Tab.B1}
\centering
\setlength{\tabcolsep}{6pt} % Default value: 6pt
\renewcommand{\arraystretch}{1.15} % Default value: 1
\begin{tabular}{@{} cccccc @{}}
\hline
$\phi$ & PSU & Estimator & AB & RMSE & $95\%$ CP  \\ \hline\hline
0.25 & 1 & $var_{\text{Coll}}$ & 9.386e-05 & 1e-04 & 0.942 \\
 && $var_{\text{Ker}}$ & 9.550e-05  & 1e-04 & 0.948 \\
 && $var_{\text{Bayes}}$ & 9.076e-05 & 1e-04 & 0.953 \\
 & 2 & $var_{\text{Coll}}$ & 2.752e-05 & 3.402e-05 & 0.955 \\
 && $var_{\text{Ker}}$ & 4.593e-05 & 5.797e-05 & 0.958 \\
 && $var_{\text{Bayes}}$ & 2.558e-05 & 3.194e-05 & 0.950 \\
\hline
0.5 & 1 & $var_{\text{Coll}}$ & 4e-04 & 5e-04 & 0.943 \\
 && $var_{\text{Ker}}$ & 4e-04 & 5e-04 & 0.948 \\
 && $var_{\text{Bayes}}$ & 3e-04 & 4e-04 & 0.950 \\
 & 2 & $var_{\text{Coll}}$ & 1.100e-04 & 1e-04 & 0.955 \\
 && $var_{\text{Ker}}$ & 1.837e-04 & 2e-04 & 0.958 \\
 && $var_{\text{Bayes}}$ & 1e-04 & 1e-04 & 0.950 \\
\hline
5 & 1 & $var_{\text{Coll}}$ & 0.038 & 0.047 & 0.942 \\
 && $var_{\text{Ker}}$ & 0.038 & 0.049 & 0.947 \\
 && $var_{\text{Bayes}}$ & 0.028 & 0.035 & 0.937 \\
 & 2 & $var_{\text{Coll}}$ & 0.011 & 0.014 & 0.955 \\
 && $var_{\text{Ker}}$ & 0.018 & 0.023 & 0.958 \\
 && $var_{\text{Bayes}}$ & 0.010 & 0.012 & 0.944 \\
\hline
\end{tabular}
\end{table}

%%%%%%%

\begin{table}[ht]
\footnotesize
\caption{Comparisons among variance estimators from the Gaussian population. Note that $b=0.0075$ for the kernel-based method and $L=7$ for the Bayesian method. We have $H=200$ strata. The results for 2 PSUs are not given here as they are computationally demanding.} \label{Tab.B2}
\centering
\setlength{\tabcolsep}{6pt} % Default value: 6pt
\renewcommand{\arraystretch}{1.15} % Default value: 1
\begin{tabular}{@{} cccccc @{}}
\hline
$\phi$ & PSU & Estimator & AB & RMSE & $95\%$ CP  \\ \hline\hline
0.25 & 1 & $var_{\text{Coll}}$ & 3.348e-05 & 4.180e-05 & 0.950 \\
 && $var_{\text{Ker}}$ & 3.337e-05 & 4.187e-05 & 0.947 \\
 && $var_{\text{Bayes}}$ & 2.708e-05 & 3.471e-05 & 0.948 \\
\hline
0.5 & 1 & $var_{\text{Coll}}$ & 1e-04 & 2e-04 & 0.950 \\
 && $var_{\text{Ker}}$ & 1e-04 & 2e-04 & 0.947 \\
 && $var_{\text{Bayes}}$ & 1e-04 & 1e-04 & 0.947 \\
\hline
5 & 1 & $var_{\text{Coll}}$ & 0.013 & 0.017 & 0.950 \\
 && $var_{\text{Ker}}$ & 0.013 & 0.017 & 0.947 \\
 && $var_{\text{Bayes}}$ & 0.009 & 0.012 & 0.943 \\
\hline
\end{tabular}
\end{table}

\begin{table}[ht]
\footnotesize
\caption{The performance of Bayesian method for different values of $L=(5,7,10)$.} \label{Tab.B3}
\centering
\setlength{\tabcolsep}{6pt} % Default value: 6pt
\renewcommand{\arraystretch}{1.15} % Default value: 1
\begin{tabular}{@{} ccccccc @{}}
\hline
$\phi$ & H & PSU & AB & RMSE & $95\%$ CP  \\ \hline\hline
0.25 & 50 & 1 & (3e-04,3e-04,3e-04) & (4e-04,4e-04,4e-04) & (0.961,0.957,0.970) \\
 & & 2 & (7.249e-05,7.159e-05,6.833e-05) & (9.184e-05,9.033e-05,8.572e-05) & (0.954,0.956,0.955) \\
 & 100 & 1 & (8.888e-05,9.076e-05,8.865e-05) & (1e-04,1e-04,1e-04) & (0.956,0.953,0.966) \\
 & & 2 & (2.481e-05,2.558e-05,2.462e-05) & (3.109e-05,3.194e-05,3.122e-05) & (0.958,0.950,0.943) \\
 & 200 & 1 & (2.879e-05,2.708e-05,2.689e-05) & (3.601e-05,3.471e-05,3.410e-05) & (0.954,0.948,0.961) \\
\hline
0.5 & 50 & 1 & (0.001,0.001,0.001) & (0.001,0.001,0.001) & (0.955,0.952,0.964) \\
 & & 2 & (3e-04,3e-04,3e-04) & (3e-04,3e-04,3e-04) & (0.951,0.956,0.955) \\
 & 100 & 1 & (3e-04,3e-04,3e-04) & (4e-04,4e-04,4e-04) & (0.954,0.950,0.958)  \\
 & & 2 & (9.703e-05,1e-04,9.168e-05) & (1e-04,1e-04,1e-04) & (0.958,0.950,0.943) \\
 & 200 & 1 & (1e-04,1e-04,1e-04) & (1e-04,1e-04,1e-04) & (0.953,0.947,0.959) \\
\hline
5 & 50 & 1 & (0.073,0.077,0.072) & (0.091,0.095,0.091) & (0.943,0.930,0.943) \\
 & & 2 & (0.025,0.026,0.024) & (0.032,0.032,0.030) & (0.942,0.945,0.947) \\
 & 100 & 1 & (0.027,0.028,0.027) & (0.034,0.035,0.035) & (0.947,0.937,0.948) \\
 & & 2 & (0.009,0.010,0.009) & (0.012,0.012,0.012) & (0.954,0.944,0.942) \\
 & 200 & 1 & (0.010,0.009,0.010) & (0.012,0.012,0.012) & (0.948,0.943,0.953) \\
\hline
\end{tabular}
\end{table}

\clearpage

\section{Details of Pseudo Likelihood} \label{App_C}
\renewcommand{\theequation}{C.\arabic{equation}}
\setcounter{equation}{0}
\renewcommand{\thelemma}{C.\arabic{lemma}}
\renewcommand{\thedefinition}{C.\arabic{definition}}

%\textbf{For the case of 1 PSU)} 
Let us assume that $\phi(\boldsymbol{y}; \mu,\sigma^2)$ is conceptually the full population likelihood. Then, the posterior density can be written as
\begin{align*}
    \pi(\boldsymbol{\Theta}, \boldsymbol{\Psi} | \boldsymbol{y}, \boldsymbol{x}) \propto \pi(\boldsymbol{\Theta}) \, \phi(\boldsymbol{y}; \mu,\sigma^2) \prod_{l=1}^{L} \phi(\beta_{q+l}; 0, \tau^2_{\beta}) \phi(\gamma_{q+l};0.\tau^2_{\gamma}),
\end{align*}
where $\boldsymbol{\Psi}=(\beta_{q+1},\ldots,\beta_{q+L}, \gamma_{q+1},\ldots,\gamma_{q+L})$.
After stratifying the population and selecting 1 PSU per stratum, and based on the observed sample $(y_h,x_h)$ for $h=1,...,H$, we modeled the data. The likelihood for the observed sample can be denoted as
\begin{equation*} \label{eq:obs_likelihood}
    \prod_{h=1}^{H} \phi(y_h; m(x_h;\beta);s^2(x_h;\gamma)).
\end{equation*}
Using the plug-in estimator for the posterior density \citep{savitsky2016bayesian}, we have the following:
\begin{align*}
    \hat{\pi}(\boldsymbol{\Theta}, \boldsymbol{\Psi} | \boldsymbol{y}, \boldsymbol{x}) \propto \pi(\boldsymbol{\Theta}) \, 
    \prod_{h=1}^{H} \phi(y_h; m(x_h;\beta);s^2(x_h;\gamma))
    \prod_{l=1}^{L} \phi(\beta_{q+l}; 0, \tau^2_{\beta}) \phi(\gamma_{q+l};0.\tau^2_{\gamma}).
\end{align*}

Then, in order to contribute the effect of each sample observation, we incorporate the sampling weight to approximate the likelihood for the population based on the sample weighted likelihood as follows
\begin{align*}
    \hat{\pi}(\boldsymbol{\Theta}, \boldsymbol{\Psi}  | \boldsymbol{y}, \boldsymbol{x}) \propto \pi(\boldsymbol{\Theta}) \, 
    \prod_{h=1}^{H} \phi(y_h; m(x_h;\beta);s^2(x_h;\gamma))^{\tilde{w}_h}
    \prod_{l=1}^{L} \phi(\beta_{q+l}; 0, \tau^2_{\beta}) \phi(\gamma_{q+l};0.\tau^2_{\gamma}),
\end{align*}
where $\tilde{w}_h$ is the normalized weight. In addition, considering $\tilde{w}_h$'s allows us to have more stabilized results compared to ignoring them.
Similarly, if we select more than 1 PSU per stratum, we can use the same strategy for ``de-correlated" observations, $\boldsymbol{z}_h=\boldsymbol{R}_h(\rho)^{-1/2}\{\boldsymbol{y}_h-m(x_h,\beta)\boldsymbol{1}_{n_h}\}$ with weight $\tilde{w}_{hj}$, as explained in Section~3.2.

\section{Effect of Pseudo Likelihood on the Results} \label{App_D}
\renewcommand{\theequation}{D.\arabic{equation}}
\setcounter{equation}{0}
\renewcommand{\thelemma}{D.\arabic{lemma}}
\renewcommand{\thedefinition}{D.\arabic{definition}}
\renewcommand{\thetable}{D.\arabic{table}}
\renewcommand{\thefigure}{D.\arabic{figure}}

In this section, we illustrate two examples based on small simulation studies, where we investigate the effect of ignoring the weights in the likelihood. We follow the simulation designs described in the manuscript and consider two scenarios of (a) Gaussian case with srswor design and selecting 1 PSU and (b) HMT case with systematic design and selecting 1 PSU.

\begin{figure}[ht]
\centering
\begin{tabular}{ c }
\includegraphics[width=0.65\textwidth]
{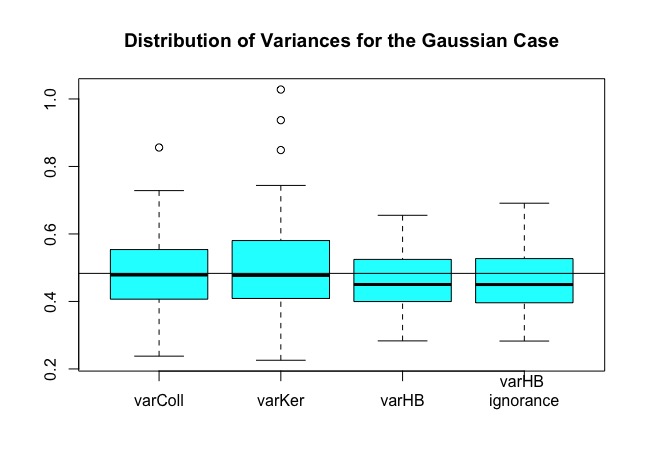} \\
\includegraphics[width=0.65\textwidth]{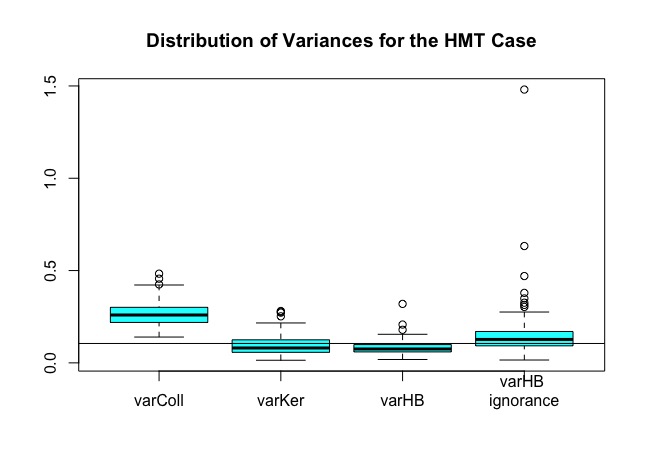} 
\end{tabular}
\caption{Comparison among the distribution of variance estimators based on 100 simulation draws. The horizontal lines are the true variance of the HT estimator $V(\bar{y}_{\text{HT}})$.}
\label{Illustration_weights}
\end{figure}

We only generate 100 draws and make comparisons among 4 variance estimators, where we ignore the weights from the likelihood for the last variance estimator of ``$var_{\text{Bayes}}$ (ignorance)". The results are summarized in Figure \ref{Illustration_weights} and Table \ref{Tab.weights}. Based on these, when we ignore the weights from the likelihood, our results become quite unstable (particularly for the HMT case with systematic design). Thus, taking into account the weights in the likelihood could provide us with some benefits, which are aligned with those given in the literature by \cite{savitsky2016bayesian} and \cite{williams2021uncertainty}.

\begin{table}[ht]
\footnotesize
\caption{Comparisons among variance estimators based on 100 simulation draws.} \label{Tab.weights}
\centering
\setlength{\tabcolsep}{6pt} % Default value: 6pt
\renewcommand{\arraystretch}{1.75} % Default value: 1
\begin{tabular}{@{} cccccc @{}}
\hline
Design & Criterion & $var_{\text{Coll}}$ & $var_{\text{Ker}}$ & $var_{\text{Bayes}}$ & $var_{\text{Bayes}}$ (ignorance) \\ \hline\hline
Gaussian & AB & 0.090 & 0.103 & 0.073 & 0.073 \\
(srswor with 1 PSU) & RMSE & 0.117 & 0.135 & 0.088 & 0.089 \\
 \hline
HMT & AB & 0.161 & 0.046 & 0.037 & 0.072 \\
(systematic with 1 PSU) & RMSE & 0.175 & 0.057 & 0.045 & 0.168 \\
\hline
\end{tabular}
\end{table}

\end{document}